\documentclass[11pt,a4paper]{article}
\pdfoutput=1

\usepackage{jheppub}
\usepackage{amsmath}
\usepackage{amssymb}
\usepackage{graphics}
\usepackage[active]{srcltx}
\usepackage{pdfsync}
\usepackage{shuffle}
\usepackage{slashed}
\usepackage{hyperref}
\usepackage{subfigure}

\newcommand{\bea}{\begin{eqnarray}}
\newcommand{\eea}{\end{eqnarray}}
\newcommand{\nn}{\nonumber \\}

\def\W #1{\widetilde{#1}}

\def\eref#1{(\ref{#1})}

\def\a{{\alpha}}

\allowdisplaybreaks

\title{Generalized $2$-split for higher-derivative YM and GR amplitudes at tree-level }
\author[a]{Liang Zhang}
\author[b]{Kang Zhou}

\affiliation[a]{Beijing Computational Science Research Center, Beijing 100084, China}
\affiliation[b]{Center for Gravitation and Cosmology, College of Physical Science and Technology, Yangzhou University,\\
No.180, Siwangting Road, Yangzhou, 225009, P.R. China}

\emailAdd{liangzhang@csrc.ac.cn}
\emailAdd{zhoukang@yzu.edu.cn}

\date{\today}
\abstract{
	We study the generalized $2$-split of higher-derivative amplitudes, including Yang-Mills (YM) and Gravity (GR) amplitudes with special insertions of higher-derivative vertices, by expanding them into ${\rm YM}\oplus{\rm BAS}$, ${\rm GR}\oplus{\rm YM}$, and ${\rm GR}\oplus{\rm YM}\oplus{\rm BAS}$ amplitude, respectively. By leveraging the established $2$-split properties of these constituent theories, we show that these higher-derivative amplitudes---which also exhibit another newly discovered phenomenon called hidden zero---do not factorize into a single product of two currents. Instead, their factorization universally appears as a sum of multiple $2$-split contributions.
}

\begin{document}

\maketitle \flushbottom

%%%%%%%%%%%%%%%%%%%%%%%%%%%%%%
\section{Introduction}
\label{sec-intro}
%%%%%%%%%%%%%%%%%%%%%%%%%%%%%%%%

Tree-level scattering amplitudes are rational functions that are fully characterized by their zeros and poles. Understanding the behavior of scattering amplitudes near their zeros and poles is therefore crucial for uncovering their underlying structures and for bootstrapping amplitudes in quantum field theory. Factorization properties around poles have long been known, including soft factorization \cite{Low:1958sn, Weinberg:1965nx} and collinear factorization \cite{Mangano:1990by}.
This property provides powerful tools for constructing tree-level amplitudes without relying on traditional Lagrangian formulations or Feynman rules. A prime example is the well-known Britto-Cachazo-Feng-Witten (BCFW) on-shell recursion relation \cite{Britto:2004ap,Britto:2005fq}, which exploits factorization on physical poles as its fundamental input. By contrast, comparatively little progress had been made in understanding the role of amplitude zeros. Recently, this gap has been partially filled by the discovery of a special class of hidden zeros in tree-level amplitudes across a wide range of theories \cite{Arkani-Hamed:2023swr}. These results show that amplitudes vanish under specific constraints on the external kinematics. As an illustrative example, consider the ${\rm Tr}(\phi^3)$ theory. One partitions the external legs into three sets, $\{i,j\}\cup A\cup B=\{1,\ldots,n\}$. The color-ordered amplitude ${\cal A}_n^{{\rm Tr}(\phi^3)}(i,\pmb A,j,\pmb B)$ then vanishes when the Mandelstam invariants satisfy $s_{ab}=0$ for all $a\in A$ and $b\in B$. Here, the bold symbols $\pmb A$ and $\pmb B$ denote ordered sets obtained by assigning specific orderings to the elements of $A$ and $B$, respectively.

Besides these hidden zeros, novel factorization behaviors associated with these zeros of amplitudes have also been discovered. One example is ``smooth splitting" \cite{Cachazo:2021wsz}, where certain scalar amplitudes split into three currents when some Mandelstam variables vanish. Another is the factorization near zeros proposed in \cite{Arkani-Hamed:2023swr}, which states that, under suitable conditions, some color-ordered amplitudes factorize into three pieces, including a four-point amplitude. An even more interesting and structurally important factorization behavior is the so-called ``$2$-split" \cite{Cao:2024gln, Cao:2024qpp}, in which amplitudes factorize on special loci in kinematic space into a product of two currents, each carrying an off-shell leg. Again taking ${\rm Tr}(\phi^3)$ theory as an example, the $2$-split occurs when a single leg $k$ is removed from the previously defined sets $A$ or $B$. Without loss of generality, let $k\in B$, such that the external legs are divided into $\{i,j,k\}\cup A\cup B'=\{1,\cdots,n\}$, where $B'\cup \{k\}=B$. Under the constraints $s_{ab}=0$ for all $a\in A$ and $b\in B'$, the amplitude factorizes as
\begin{align}
{\cal A}^{{\rm Tr}(\phi^3)}_n(i,\pmb A,j,\pmb B'(k))\to
{\cal J}^{{\rm Tr}(\phi^3)}(i,\pmb A,j,\kappa)\times {\cal J}^{{\rm Tr}(\phi^3)}(j,\pmb B',i;\kappa')\,,~~\label{split-phi3}
\end{align}
where ${\cal J}^{{\rm Tr}(\phi^3)}(i,\pmb A,j,\kappa)$ and ${\cal J}^{{\rm Tr}(\phi^3)}(j,\pmb B',i;\kappa')$ are amputated currents carrying the off-shell external legs $\kappa$ and $\kappa'$, respectively. The notation $\pmb B'(k)$ denotes the ordered set where the leg $k$ is inserted within the set $\pmb B'$. Remarkably, the special leg $k$ effectively transmutes into the off-shell legs $\kappa$ and $\kappa'$ in the two resulting currents.

Remarkably, the former two factorization behaviors can be recovered from the $2$-split by imposing additional constraints. Owing to its generality and significance, the $2$-split has attracted considerable attention and has been investigated using a variety of methods and in different theories (see, e.g., \cite{Arkani-Hamed:2024fyd, Zhou:2024ddy, Zhang:2024efe, Zhang:2024iun, Feng:2025ofq, Feng:2025dci, Zhang:2025zjx, Azevedo:2025vxo}).

In literature \cite{Arkani-Hamed:2023swr}, theories known to exhibit hidden zeros---including ${\rm Tr}(\phi^3)$, the Nonlinear sigma model (NLSM), Yang-Mills (YM), General Relativity (GR), Dirac-Born-Infeld (DBI), and the Special Galileon (SG)---consistently exhibit the $2$-split behavior simultaneously, and vice versa. Given the striking similarity between the kinematic conditions required for hidden zeros and those for the $2$-split, it is natural to expect a close connection between these two significant behaviors. While this relationship has been investigated from various perspectives \cite{Cao:2024gln, Cao:2024qpp, Feng:2025dci, Feng:2025ofq}, a definitive physical principle remains elusive to guarantee that the behavior of amplitudes with hidden zeros must necessarily follow the specific structure of \eqref{split-phi3}. This raises a question: Does the presence of a hidden zero always imply the $2$-split as defined in \eqref{split-phi3}, or must this factorization behavior be generalized when considering a broader range of theories?

To answer this question, in this paper we study the amplitudes of YM and general GR theories supplemented by gauge invariant higher-derivative operators, whose hidden zeros have recently been identified in \cite{Zhou:2025tvq}. In particular, we focus on the amplitude ${\cal A}_n^{F^3}$, which describes $n$ gluons with a single insertion of the $F^3$ vertex,
\begin{align}
F^3 \equiv {\rm Tr}(F_{\mu_1}^{\nu_1}F_{\mu_2}^{\nu_2}F_{\mu_3}^{\nu_3})
= \frac{1}{2}f^{abc}F_{\mu_1}^{a\nu_1}F_{\mu_2}^{b\nu_2}F_{\mu_3}^{c\nu_3}\,,
\end{align}
where $F_{\mu\nu}\equiv F^a_{\mu\nu}T^a$ is the gluon field strength and $f^{abc}={\rm Tr}([T^a,T^b]T^c)$ are the structure constants of the gauge group. This operator arises as the sub-leading correction to the YM Lagrangian in the $\alpha'$ expansion of bosonic open-string theory \cite{Polchinski:1998rq}. We further consider the $R^2$ amplitude ${\cal A}_n^{R^2}$ as the double copy of ${\cal A}_n^{\rm YM}$ with ${\cal A}_n^{F^3}$, and the $R^3$ amplitude ${\cal A}_n^{R^3}$ as the double copy of ${\cal A}_n^{F^3}$ with itself \cite{He:2016iqi}. As will be demonstrated, these amplitudes can be interpreted as graviton amplitudes with the insertion of higher-derivative interactions. By imposing kinematic conditions identical to those required for the $2$-split of standard YM and GR amplitudes, we investigate the factorization behaviors of these new amplitudes in the presence of higher-derivative interactions.

Our approach is based on expanding these higher-derivative amplitudes into amplitudes in simpler constituent theories, namely ${\rm BAS}\oplus{\rm YM}$, ${\rm GR}\oplus{\rm YM}$, and ${\rm GR}\oplus{\rm YM}\oplus{\rm BAS}$, where BAS denotes the bi-adjoint scalar. While the $2$-split of conventional YM and GR amplitudes is well-established, the $2$-splits of these mixed amplitudes can be systematically derived by applying the transmutation operators introduced in \cite{Cheung:2017ems}. By exploiting the known $2$-split properties of these constituent theories, and carefully treating the expansion coefficients, we demonstrate that higher-derivative amplitudes do not factorize into a single $2$-split term. Instead, their factorization structure generically takes the form of a sum of $2$-splits:
\bea
{\cal A}_n\,\to\,\sum_{i}\,{\cal J}_{i1}\,\times\,{\cal J}_{i2}\,.~~\label{split-generalized}
\eea
We also show that such behavior is consistent with $2$-split structures of full bosonic string amplitudes given by \cite{Cao:2024gln, Cao:2024qpp}
\bea
{\cal M}^{\rm open}\,\to\,{\cal J}^{\rm open}_1\,\times\,{\cal J}^{\rm open}_2\,,~~{\cal M}^{\rm closed}\,\to\,{\cal J}^{\rm closed}_1\,\times\,{\cal J}^{\rm closed}_2\,.~~\label{split-fullstring}
\eea

The remainder of this paper is organized as follows. Section~\ref{sec-fac-F3} presents the full factorization of $F^3$ amplitudes, analyzing the three sectors of admissible subsets contributing to the expansion index $\rho$, which play a central role throughout this work. Section~\ref{sec-R2R3} extends the analysis to $R^2$ and $R^3$ amplitudes using similar methods. Finally, we summarize our results and discuss future directions.

\paragraph{Note.}
While finalizing this paper, we became aware of \cite{Azevedo:2025vxo}, where $2$-split factorizations were also reported for theories denoted as $R^2$ and $R^3$. Despite the similar nomenclature, these theories are not identical to those studied here. In \cite{Azevedo:2025vxo}, the $R^2$ and $R^3$ theories arise as double copies of $(DF)^2$ with YM and with $(DF)^2$, respectively \cite{Johansson:2017srf}, whereas in our work they are constructed as double copies of $F^3$ with YM and with $F^3$, respectively.

%%%%%%%%%%%%%%%%%%%%%%%%%%%%%%
\section{Factorization of $F^3$ amplitudes}
\label{sec-fac-F3}
%%%%%%%%%%%%%%%%%%%%%%%%%%%%%%

In this section, we investigate the $2$-split behavior of $F^3$ amplitudes. As noted in the introduction, these amplitudes represent the sub-leading contribution to bosonic open-string amplitudes. Specifically, when the open-string amplitude is expanded in powers of $\a'$, the  interaction corresponds to the $\mathcal{O}(\alpha')$ term. The CHY integrand for these higher-derivative amplitudes is given by:
\begin{align}
{\cal I}^{ F^3}_n(\pmb\sigma_n)={\cal P}_n{\rm PT}(\pmb\sigma_n)\,, \label{CHY-F3}
\end{align}
where $\pmb\sigma_n$ denotes the color ordering of  external gluons and ${\rm PT}(\pmb\sigma_n)$ is the associated Parke-Taylor factor for $n$ punctures. The explicit construction of the building block ${\cal P}_n$ can be found in \cite{He:2016iqi}.

We utilize the expansion formula for higher-derivative amplitudes established in \cite{Bonnefoy:2023imz,Zhou:2024qwm}:
\bea
{\cal A}^{F^3}_n(\pmb\sigma_n)=\sum_{\pmb\rho,2\leq|\rho|\leq n}{\rm Tr}({\rm F}_{\pmb\rho}){\cal A}^{{\rm BAS}\oplus{\rm YM}}_n(\pmb\rho|\pmb\sigma_n)\,,
\eea
where the summation runs over all cyclically inequivalent ordered sets $\pmb\rho$ with cardinality $2\leq|\rho|\leq n$. From the CHY perspective, this expansion can be understood as
\bea
{\cal P}_n=\sum_{\pmb\rho,2\leq|\rho|\leq n}{\rm Tr}({\rm F}_{\pmb\rho})\,{\rm PT}(\pmb\rho)\,{\rm Pf}'\Psi^{\overline{\rho}}\,,~~\label{expan-P}
\eea
where ${\rm Pf}'\Psi^{\overline{\rho}}$ denotes the reduced pfaffian for $\{1,\cdots,n\}\setminus\rho$. Here we have used the observation that the expansion does not affect the measure $d\mu_n$ in the CHY formula
\bea
{\cal A}_n=\int\,d\mu_n\,{\cal I}^{\rm CHY}\,,
\eea
which implies that the expansion of amplitude is equivalent to the expansion of CHY integrand ${\cal I}^{\rm CHY}$.

We focus on $F^3$ amplitudes with the special color orderings $(i,\pmb A,j,\pmb B(k))$, where the expansion is given by:
\begin{align}
{\cal A}^{F^3}_n(i,\pmb A,j,\pmb B(k))=\sum_{\pmb\rho,2\leq|\rho|\leq n}{\rm Tr}({\rm F}_{\pmb\rho}){\cal A}^{{\rm BAS}\oplus{\rm YM}}_n(\pmb\rho|i,\pmb A,j,\pmb B(k))\,,
\label{expan-F3}\end{align}
where $\pmb B(k)$ can be identified with $\pmb B'(k)$ in \eref{split-phi3}. Throughout this paper, we adopt the convention of using ordinary Greek letters and uppercase Latin letters for unordered sets, while bold symbols denote their ordered counterparts3. For an ordered set $\pmb\rho=\{\rho_1,\rho_2,\ldots,\rho_{|\rho|}\}$, the corresponding kinematic factor is defined as:
\begin{align}{\rm Tr}({\rm F}_{\pmb\rho})\equiv (-1)^{|\rho|}\left(f_{\rho_1}\cdot f_{\rho_2}\cdot\ldots \cdot f_{\rho_{|\rho|}}\right)_{\mu}^{~\mu}\,,\qquad \text{with} \qquad f_i^{\mu\nu}=k_i^\mu \epsilon_i^\nu - \epsilon_i^\mu k_i^\nu\,.
\end{align}

We impose the following condition on external kinematics,
\begin{align}
\{\epsilon_{i,j,k},\epsilon_a,k_a\}\cdot \{\epsilon_b,k_b\}=0\,,    \qquad \forall a\in A, b\in B\,.\label{kinematic-condition}
\end{align}
This condition yields the $2$-split of standard YM amplitudes \cite{Cao:2024qpp}:
\bea
{\cal A}^{\rm YM}_n(i,\pmb A,j,\pmb B(k))
\xrightarrow[]{\eref{kinematic-condition}}
{\cal J}^{\rm YM}_{|A|+3}(i,\pmb A,j,\kappa)\times
{\cal J}^{{\rm Tr}(\phi^3)\oplus{\rm YM}}_{|B|+3}(i_\phi,j_\phi,\pmb B(\kappa'_\phi))\,.\label{2split-YM}
\eea
As we will demonstrate, under the same kinematic condition, the $F^3$ amplitude exhibits a factorization behavior as describe in \eref{split-generalized}.

Under the constraints \eref{kinematic-condition}, the kinematic factors ${\rm Tr}({\rm F}_{\pmb\rho})$ vanish for certain ordered sets $\pmb\rho$, causing them to drop out of the expansion \eqref{expan-F3}. Consequently, any admissible set $\pmb\rho$ must satisfy the following criteria:
\begin{itemize}\item[(r1)] Elements from $A$ cannot be adjacent to elements from $B$, as the contraction $\{\epsilon_a,k_a\}\cdot \{\epsilon_b,k_b\}$ vanishes under \eqref{kinematic-condition}.
\item[(r2)] No element $c\in\{i,j,k\}$ can be inserted between two elements from $B$. Such a configuration would force $\epsilon_c$ to contract with either $k_b$ or $\epsilon_b$, both of which are zero according to \eqref{kinematic-condition} .\footnote{By the same logic, configurations where $\pmb\rho$ contains an element $c$ adjacent to only a single element from $B$ are also excluded.}
\end{itemize}
Based on these requirements, the admissible sets $\pmb\rho$ in \eqref{expan-F3} are naturally partitioned into three distinct sectors:
\begin{itemize}
\item[(1)] $\pmb\rho$ is a non-empty subset of $A_{ijk}\equiv A\cup\{i,j,k\}$, denoted by $\rho = A_{ijk}^{\rm sub}$.
\item[(2)] $\pmb\rho$ is a non-empty subset of $B$, denoted $\rho = B^{\rm sub}$.
\item[(3)] $\pmb\rho$ contains elements from both $B$ and $\{i,j,k\}$, potentially including elements from $A$. Requirement (r2) dictates that elements of $B$ must be nested between two elements $\{m,n\}\subset\{i,j,k\}$. Combined with (r1), the only admissible ordered sets in this sector take the form $\{m,{\pmb D},n,\pmb B^{\rm sub}\}$, where $D$ is a (possibly empty) subset of $A_l=A\cup\{i,j,k\}/\{m,n\}$. We denote these sets as $\rho = \{m,n\}\cup D\cup B^{\rm sub}$.
\end{itemize}

%%%%%%%%%%%%%%%%%%%%%%%%%%%%
\subsection{Case (1) : $\rho = A_{ijk}^{\rm sub}$}
\label{subsec-F3-part1}
%%%%%%%%%%%%%%%%%%%%%%%%%%%%

In this subsection, we consider the first case $\rho = A_{ijk}^{\rm sub}$, where $\pmb\rho$ contains no elements from the set $B$. For this configuration, the amplitudes exhibit the following factorization behavior under the kinematic constraints \eqref{kinematic-condition}:
\bea
&&{\cal A}^{{\rm BAS}\oplus{\rm YM}}_n(\pmb\rho|i,\pmb A,j,\pmb B(k))
\xrightarrow[]{\eref{kinematic-condition}}
{\cal J}^{{\rm BAS}\oplus{\rm YM}}_{|A|+3}(\pmb\rho|i,\pmb A,j,\kappa)\times
{\cal J}^{{\rm Tr}(\phi^3)\oplus{\rm YM}}_{|B|+3}(i_\phi,j_\phi,\pmb B(\kappa'_\phi))\,,\label{2split-YMS}
\eea
where $\kappa$ and $\kappa'$ denote the resulting off-shell legs. While this $2$-split can be derived via established expansion methods and Feynman diagrammatic analysis, a more direct derivation utilizes the transmutation operators that map pure YM amplitudes to BAS$\oplus$YM amplitudes \cite{Cheung:2017ems}
\bea
{\cal A}^{{\rm BAS}\oplus{\rm YM}}_n(\pmb\rho|i,\pmb A,j,\pmb B(k))
={\cal T}[\pmb\rho]
{\cal A}^{\rm YM}_n(i,\pmb A,j,\pmb B(k))\,,
\eea
where the composite operator is defined as:
\begin{align}
	{\cal T}[\pmb\rho]
	\equiv
	{\cal T}_{\rho_1\rho_{|\rho|}}
	\prod_{i=2}^{|\rho|-1}{\cal T}_{\rho_{i-1}\rho_i\rho_{|\rho|}}\,.
\end{align}
Here, we define the trace operator ${\cal T}_{ij}=\partial_{\epsilon_i\cdot\epsilon_j}$ and the insertion operator ${\cal T}_{ijk}=\partial_{k_i\cdot\epsilon_j}-\partial_{k_k\cdot\epsilon_j}$.

Briefly reviewing these operators: the trace operator ${\cal T}_{ij}$ reduces the spin of particles $i$ and $j$ by one unit, placing them within a new color trace structure. The insertion operator ${\cal T}_{ijk}$ reduces the spin of particle $j$ and inserts it between particles $i$ and $k$ within an existing trace. Consequently, ${\cal T}[\pmb\rho]$  transmutes gravitons/gluons in the set $\pmb\rho$ into gluons/BASs forming a new color trace structure.

As established in \cite{Cao:2024gln, Cao:2024qpp}, pure YM amplitudes factorize as:
For case (1), the operator ${\cal T}[\pmb\rho]$ acts exclusively on the first current ${\cal J}^{\rm YM}_{|A|+3}(i,\pmb A,j,\kappa)$ in \eref{2split-YM}. Thus, the transmutation operator ${\cal T}[\pmb\rho]$ maps the YM $2$-split directly into the BAS$\oplus$YM $2$-split presented in \eqref{2split-YMS}.

By substituting the $2$-split relation \eqref{2split-YMS} into the expansion formula \eqref{expan-F3}, we obtain the factorization for the contribution in case (1):
\bea
P_{(1)}&\xrightarrow[]{\eref{kinematic-condition}}&
\sum_{\pmb\rho\,, \rho= A_{ijk}^{\rm sub}}
{\rm Tr}({\rm F}_{\pmb\rho})
{\cal J}^{{\rm BAS}\oplus{\rm YM}}_{|A|+3}(\pmb\rho|i,\pmb A,j,\kappa)\times
{\cal J}^{{\rm Tr}(\phi^3)\oplus{\rm YM}}_{|B|+3}(i_\phi,j_\phi,\pmb B(\kappa'_\phi))\nn
&=&
\epsilon_k\cdot{\cal J}^{F^3}_{|A|+3}(i,\pmb A,j,\kappa)\times
{\cal J}^{{\rm Tr}(\phi^3)\oplus{\rm YM}}_{|B|+3}(i_\phi,j_\phi,\pmb B(\kappa'_\phi))\,,\label{2split-part1}
\eea
where the $F^3$ current is defined as:
\bea
\epsilon_k\cdot{\cal J}^{F^3}_{|A|+3}(i,\pmb A,j,\kappa)=
\sum_{\pmb\rho,\ 2\leq|\rho|\leq |A|+3}
{\rm Tr}({\rm F}_{\pmb\rho})\big|_{k_\kappa\to k_k}
{\cal J}^{{\rm BAS}\oplus{\rm YM}}_{|A|+3}(\pmb\rho|i,\pmb A,j,\kappa)\,.\label{J-F3}
\eea
Note that in the definition \eqref{J-F3}, the field strength associated with the off-shell leg is interpreted as $f_{\kappa}^{\mu\nu} = k_k^\mu \epsilon_k^\nu - \epsilon_k^\mu k_k^\nu$ (using the momentum of the external leg $k$) rather than $k_\kappa^\mu \epsilon_k^\nu - \epsilon_k^\mu k_\kappa^\nu$.

%%%%%%%%%%%%%%%%%%%%%%%
\subsection{Case (2) and (3) : $\rho=B^{\rm sub}$ \& $\rho = \{m,n\}\cup D\cup B^{\rm sub}$}
\label{subsec-F3-part2}
%%%%%%%%%%%%%%%%%%%%%%%
\textbf{Case (2)}: $\rho = B^{\rm sub}$\\
Next, we consider the second case where $\pmb \rho$ is a non-empty subset of $B$ ($\pmb \rho = B^{\rm sub}$). In this configuration, the ${\rm BAS}\oplus{\rm YM}$ amplitudes factorize as follows:
\bea
&&{\cal A}^{{\rm BAS}\oplus{\rm YM}}_n(\pmb B^{\rm sub}|i,\pmb A,j,\pmb B(k))
\xrightarrow[]{\eref{kinematic-condition}}
{\cal J}^{\rm YM}_{|A|+3}(i,\pmb A,j,\kappa)\times
{\cal J}^{{\rm BAS}\oplus{\rm YM}}_{|B|+3}(\pmb B^{\rm sub};i,j,k|i_\phi,j_\phi,\pmb B(\kappa'_\phi))\,.
\label{2split-doubletr}
\eea
This behavior arises because the transmutation operator ${\cal T}[\pmb B^{\rm sub}]$, which maps the pure YM amplitude ${\cal A}^{\rm YM}_n$ to the ${\rm BAS}\oplus{\rm YM}$ amplitude, acts exclusively on the second factor, the current ${\cal J}^{{\rm Tr}(\phi^3)\oplus{\rm YM}}_{|B|+3}$, within the YM $2$-split given in \eqref{2split-YM}.

The resulting current is a double-trace ${\rm BAS}\oplus{\rm YM}$ current, where the two traces are defined by the orderings $\pmb B^{\rm sub}$ and $(i,j,k)$. To derive this, we utilize the relation:
\begin{align}
{\cal J}^{{\rm Tr}(\phi^3)\oplus{\rm YM}}_{|B|+3}(i_\phi,j_\phi,\pmb B(\kappa'_\phi)) = {\cal J}^{{\rm BAS}\oplus{\rm YM}}_{|B|+3}(i,j,\kappa'|i,j,\pmb B(\kappa'))\,, \label{phi3-BAS}
\end{align}
which identifies the ${\rm Tr}(\phi^3)\oplus{\rm YM}$ current as a specific case of a ${\rm BAS}\oplus{\rm YM}$ current. By substituting the factorization relation \eqref{2split-doubletr} into the expansion formula \eqref{expan-F3}, we obtain the contribution for case (2):
\bea
P_{(2)}
\xrightarrow[]{\eref{kinematic-condition}}
{\cal J}^{\rm YM}_{|A|+3}(i,\pmb A,j,\kappa)\times
\Bigg(\sum_{\pmb B^{\rm sub}}
{\rm Tr}({\rm F}_{\pmb B^{\rm sub}})
{\cal J}^{{\rm BAS}\oplus{\rm YM}}_{|B|+3}(\pmb B^{\rm sub};i,j,k|i_\phi,j_\phi,\pmb B(\kappa'_\phi))\Bigg)\,.
\label{2split-case6}
\eea
\textbf{Case (3)}: $\rho = \{m,n\}\cup D\cup B^{\rm sub}$\\
Finally, we examine the third case, where $\rho = \{m,n\}\cup D\cup B^{\rm sub}$. Recall that the ordered set $\pmb\rho$ is
%The kinematic condition \eref{kinematic-condition} implies that the non-vanishing ${\rm Tr}({\rm F}_{\pmb\rho})$ corresponds to
%%
\bea
 \pmb\rho=\{m, \pmb D, n,\pmb B^{\rm sub}\}\,.~~~~\label{rhoAB}
\eea
For any $\pmb\rho$ satisfying \eqref{rhoAB}, the associated kinematic factor factorizes as:
\bea
&&{\rm Tr}({\rm F}_{\pmb\rho})\xrightarrow[]{\eref{kinematic-condition}}\Big((-)^{|D|}\epsilon_m\cdot {\rm F}_{\pmb D}\cdot\epsilon_n\Big)\times
\Big((-)^{|B^{\rm sub}|-1}k_n\cdot {\rm F}_{\pmb B^{\rm sub}}\cdot k_m\Big)\,,~~\label{fac-trF}
\eea
where  the tensors ${\rm F}_{\pmb D}$ and ${\rm F}_{\pmb B^{\rm sub}}$ are defined by the product of field strengths:
\bea
{\rm F}_{\pmb D}^{\mu\nu}\equiv \big(f_{d_1}\cdot f_{d_2}\cdot\ldots\cdot f_{d_{|D|}}\big)^{\mu\nu}\,,~~~~~~
{\rm F}_{\pmb B^{\rm sub}}^{\mu\nu}\equiv \big(f_{b_1}\cdot f_{b_2}\cdot\ldots\cdot f_{b_{|B^{\rm sub}|}}\big)^{\mu\nu}\,.
\eea
Here,
\bea
\pmb D=\{d_1,d_2,\ldots,d_{|D|}\}\,,~~~~~\pmb B^{\rm sub}=\{b_1,b_2,\ldots,b_{|B^{\rm sub}|}\}\,.
\eea
Note that the set $D$ may be empty. Substituting this factorization \eqref{fac-trF} into the expansion formula \eqref{expan-F3} yields the contribution for case (3):
\bea
P_{(3)}&\xrightarrow[]{\eref{kinematic-condition}}&\sum_{(m,n)}\sum_{\pmb D}\sum_{\pmb B^{\rm sub}}\Big((-)^{|D|}\epsilon_m\cdot {\rm F}_{\pmb D}\cdot\epsilon_n\Big)\times
\Big((-)^{|B^{\rm sub}|-1}k_n\cdot {\rm F}_{\pmb B^{\rm sub}}\cdot k_m\Big)\nn
&&~~~~~~~~~~~~~~~~~~{\cal A}^{{\rm BAS}\oplus{\rm YM}}_n(m,\pmb D,n,\pmb B^{\rm sub}|i,\pmb A,j,\pmb B(k))\,.~~{\label{part2-step1}}
\eea

The BAS$\oplus$YM amplitudes in \eqref{part2-step1} further exhibit a $2$-split behavior:
\bea
&&{\cal A}^{{\rm BAS}\oplus{\rm YM}}_n(m,\pmb D,n,\pmb B^{\rm sub}|i,\pmb A,j,\pmb B(k))\xrightarrow[]{\eref{kinematic-condition}}\nn
&&~~~~~~~~{\cal J}^{{\rm BAS}\oplus{\rm YM}}_{|A|+3}(m,\pmb D,n|i,\pmb A,j,\kappa)\times{\cal J}^{{\rm BAS}\oplus{\rm YM}}_{|B|+3}(\pmb B^{\rm sub}_{[ijk]}|i,j,\pmb B(\kappa'))\,,~~\label{2split-YMS2}
\eea
which is derived by applying the differential operators to the YM factorization in \eqref{2split-YM}. Specifically, the operator ${\cal T}[\pmb\rho]$ factorizes under the kinematic constraints as:
\bea
{\cal T}[\pmb\rho]\xrightarrow[]{\eref{kinematic-condition}}\Big(\partial_{\epsilon_m\cdot\epsilon_n}\prod_{p=1}^{|D|}(\partial_{\epsilon_{d_p}\cdot k_{d_{p-1}}}-\partial_{\epsilon_{d_p}\cdot k_n})\Big)\times
\Big(\prod_{q=1}^{|B^{\rm sub}|}(\partial_{\epsilon_{b_q}\cdot k_{b_{q-1}}}-\partial_{\epsilon_{b_q}\cdot k_m})\Big)\,,~~\label{operator}
\eea
where $k_{d_{0}}=k_m$ and $k_{b_{0}}=k_n$. The first operator factor acts solely on ${\cal J}^{\rm YM}_{|A|+3}$, transmuting it into the current ${\cal J}^{{\rm BAS}\oplus{\rm YM}}_{|A|+3}(m,\pmb D,n|i,\pmb A,j,\kappa)$, when $k\not\in D$, this current should be interpreted as
\bea
{\cal J}^{{\rm BAS}\oplus{\rm YM}}_{|A|+3}(m,\pmb D,n|i,\pmb A,j,\kappa)\sim\epsilon_k\cdot{\cal J}^{{\rm BAS}\oplus{\rm YM}}_{|A|+3}(m,\pmb D,n|i,\pmb A,j,\kappa)\,.~~\label{k-notin-D}
\eea
While the second factor acts on ${\cal J}^{{\rm Tr}(\phi^3)\oplus{\rm YM}}_{|B|+3}$ to insert the set $\pmb B^{\rm sub}$ between $n$ and $m$. The resulting configurations for the second current, depending on the choice of $(n, m) \subset \{i, j, \kappa'\}$, are summarized in \eqref{Bijk}:
\bea
&&{\cal J}^{{\rm BAS}\oplus{\rm YM}}_{|B|+3}(\pmb B^{\rm sub}_{[ijk]}|i,j,\pmb B(\kappa'))={\cal J}^{{\rm BAS}\oplus{\rm YM}}_{|B|+3}(i,\pmb B^{\rm sub},j,\kappa'|i,j,\pmb B(\kappa'))\,,~~~~{\rm if}~(n,m)=(i,j)\,,\nn
&&{\cal J}^{{\rm BAS}\oplus{\rm YM}}_{|B|+3}(\pmb B^{\rm sub}_{[ijk]}|i,j,\pmb B(\kappa'))=(-)^{|B^{\rm sub}|}{\cal J}^{{\rm BAS}\oplus{\rm YM}}_{|B|+3}(i,\pmb B^{{\rm sub};T},j,\kappa'|i,j,\pmb B(\kappa'))\,,~~~~{\rm if}~(n,m)=(j,i)\,,\nn
&&{\cal J}^{{\rm BAS}\oplus{\rm YM}}_{|B|+3}(\pmb B^{\rm sub}_{[ijk]}|i,j,\pmb B(\kappa'))={\cal J}^{{\rm BAS}\oplus{\rm YM}}_{|B|+3}(i,j,\pmb B^{\rm sub},\kappa'|i,j,\pmb B(\kappa'))\,,~~~~{\rm if}~(n,m)=(j,\kappa')\,,\nn
&&{\cal J}^{{\rm BAS}\oplus{\rm YM}}_{|B|+3}(\pmb B^{\rm sub}_{[ijk]}|i,j,\pmb B(\kappa'))=(-)^{|B^{\rm sub}|}{\cal J}^{{\rm BAS}\oplus{\rm YM}}_{|B|+3}(i,j,\pmb B^{{\rm sub};T},\kappa'|i,j,\pmb B(\kappa'))\,,~~~~{\rm if}~(n,m)=(\kappa',j)\,,\nn
&&{\cal J}^{{\rm BAS}\oplus{\rm YM}}_{|B|+3}(\pmb B^{\rm sub}_{[ijk]}|i,j,\pmb B(\kappa'))=(-)^{|B^{\rm sub}|}{\cal J}^{{\rm BAS}\oplus{\rm YM}}_{|B|+3}(i,j,\kappa',\pmb B^{{\rm sub};T}|i,j,\pmb B(\kappa'))\,,~~~~{\rm if}~(n,m)=(i,\kappa')\,,\nn
&&{\cal J}^{{\rm BAS}\oplus{\rm YM}}_{|B|+3}(\pmb B^{\rm sub}_{[ijk]}|i,j,\pmb B(\kappa'))={\cal J}^{{\rm BAS}\oplus{\rm YM}}_{|B|+3}(i,j,\kappa',\pmb B^{\rm sub}|i,j,\pmb B(\kappa'))\,,~~~~{\rm if}~(n,m)=(\kappa',i)\,,
\label{Bijk}
\eea
where the relation \eref{phi3-BAS} has been used.

By inserting the $2$-split relation \eqref{2split-YMS2} into \eqref{part2-step1}, we arrive at:
\bea
P_{(3)}&\xrightarrow[]{\eref{kinematic-condition}}&\sum_{(m,n)}\Big(\sum_{\pmb D}(-)^{|D|}(\epsilon_m\cdot {\rm F}_{\pmb D}\cdot\epsilon_n){\cal J}^{{\rm BAS}\oplus{\rm YM}}_{|A|+3}(m,\pmb D,n|i,\pmb A,j,\kappa)\Big)\nn
&&\times\Big(\sum_{\pmb B^{\rm sub}}(-)^{|B^{\rm sub}|-1}(k_n\cdot {\rm F}_{\pmb B^{\rm sub}}\cdot k_m){\cal J}^{{\rm BAS}\oplus{\rm YM}}_{|B|+3}(\pmb B^{\rm sub}_{[ijk]}|i,j,\pmb B(\kappa'))\Big)\,.\label{2split-case7}
\eea
Crucially, the first summation in \eqref{2split-case7} can be identified as the YM current for any pair $(m, n)$:
\bea
\sum_{\pmb D}(-)^{|D|}(\epsilon_m\cdot {\rm F}_{\pmb D}\cdot\epsilon_n){\cal J}^{{\rm BAS}\oplus{\rm YM}}_{|A|+3}(m,\pmb D,n|i,\pmb A,j,\kappa)
=\epsilon_k\cdot{\cal J}^{\rm YM}_{|A|+3}(i,\pmb A,j,\kappa)\,,~~\label{recognize}
\eea
This identity is verified by expanding the pure YM amplitude as
\bea
{\cal A}^{\rm YM}_n(i,\pmb A,j,\pmb B(k))=\sum_{\pmb D'}(-)^{|D'|}(\epsilon_m\cdot{\rm F}_{\pmb D'}\cdot\epsilon_n){\cal A}^{{\rm BAS}\oplus{\rm YM}}_n(m,\pmb D',n|i,\pmb A,j,\pmb B(k))\,.~~\label{expan-forYM}
\eea
The kinematic constraints \eqref{kinematic-condition} restrict the expansion to sets $D'=D$ that contain no elements of $B$. By comparing the resulting expanded factorization with \eqref{2split-YM}, the relation \eqref{recognize} follows immediately. Each BAS$\oplus$YM amplitude factorizes as
\bea
{\cal A}^{{\rm BAS}\oplus{\rm YM}}_n(m,\pmb D,n|i,\pmb A,j,\pmb B(k))\xrightarrow[]{\eref{kinematic-condition}}{\cal J}^{{\rm BAS}\oplus{\rm YM}}_{|A|+3}(m,\pmb D,n|i,\pmb A,j,\kappa)\times{\cal J}^{{\rm Tr}(\phi^3)\oplus{\rm YM}}_{|B|+3}(i_{\phi},j_\phi,\pmb B(\kappa'_\phi))\,,~~\label{fac-forYM}
\eea
because ${\cal T}[m,\pmb D,n]$ acts only on ${\cal J}^{\rm YM}_{|A|+3}$ in \eref{2split-YM}.
Substituting \eref{fac-forYM} into \eref{expan-forYM} and using $D'=D$ yields
\bea
{\cal A}^{\rm YM}_n(i,\pmb A,j,\pmb B(k))\xrightarrow[]{\eref{kinematic-condition}}&&\Big(\sum_{\pmb D}(-)^{|D|}(\epsilon_m\cdot {\rm F}_{\pmb D}\cdot\epsilon_n){\cal J}^{{\rm BAS}\oplus{\rm YM}}_{|A|+3}(m,\pmb D,n|i,\pmb A,j,\kappa)\Big)\nn
&&\times{\cal J}^{{\rm Tr}(\phi^3)\oplus{\rm YM}}_{|B|+3}(i_{\phi},j_\phi,\pmb B(\kappa'_\phi))\,,\label{2split-YM-expanded}
\eea
and comparison with \eref{2split-YM} immediately yields \eref{recognize}.

Finally, by combining the results from \eref{2split-case6} and \eref{2split-case7}, we obtain the unified $2$-split contribution:
\bea
P_{(2),(3)}\xrightarrow[]{\eref{kinematic-condition}}\epsilon_k\cdot{\cal J}^{\rm YM}_{|A|+3}(i,\pmb A,j,\kappa)\times{\cal J}^X_{|B|+3}(i_\phi,j_\phi,\pmb B(\kappa'_\phi))\,,~~\label{2split-part2}
\eea
where the composite current ${\cal J}^X$ is defined as:
\bea
{\cal J}^X_{|B|+3}(i_\phi,j_\phi,\pmb B(\kappa'_\phi))&=&\sum_{\pmb B^{\rm sub}}{\rm Tr}({\rm F}_{\pmb B^{\rm sub}}){\cal J}^{{\rm BAS}\oplus{\rm YM}}_{|B|+3}(\pmb B^{\rm sub};i,j,k|i_\phi,j_\phi,\pmb B(\kappa'_\phi))\nn
&&+\sum_{(m,n)}\sum_{\pmb B^{\rm sub}}(-)^{|B^{\rm sub}|-1}(k_n\cdot {\rm F}_{\pmb B^{\rm sub}}\cdot k_m){\cal J}^{{\rm BAS}\oplus{\rm YM}}_{|B|+3}(\pmb B^{\rm sub}_{[ijk]}|i,j,\pmb B(\kappa'))\,.\label{JX}
\eea

%%%%%%%%%%%%%%%%%%%%%%%%
\subsection{The full factorization}
\label{subsec-F3-full}
%%%%%%%%%%%%%%%%%%%%%%%%

By combining the partial results from \eqref{2split-part1} and \eqref{2split-part2}, we arrive at the complete factorization formula for the $F^3$ amplitude:
\bea
{\cal A}^{F^3}_n(i,\pmb A,j,\pmb B(k))&\xrightarrow[]{\eref{kinematic-condition}}&\epsilon_k\cdot{\cal J}^{F^3}_{|A|+3}(i,\pmb A,j,\kappa)\times
{\cal J}^{{\rm Tr}(\phi^3)\oplus{\rm YM}}_{|B|+3}(i_\phi,j_\phi,\pmb B(\kappa'_\phi))\nn
&&+\epsilon_k\cdot{\cal J}^{\rm YM}_{|A|+3}(i,\pmb A,j,\kappa)\times{\cal J}^X_{|B|+3}(i_\phi,j_\phi,\pmb B(\kappa'_\phi))\,.~~\label{split-F3}
\eea
The full amplitude naturally decomposes into two contributions, each of which exhibits a distinct $2$-split structure.

This structural decomposition is physically intuitive. From a field-theoretic perspective, a standard YM amplitude factorizes into two currents. When a local $F^3$ operator is inserted into the theory, the insertion may be absorbed by either of the two resulting currents. These two distinct possibilities necessitate that the full amplitude be expressed as a sum of these separate parts.

This behavior is further supported by string-theoretic considerations. The full bosonic open-string amplitude ${\cal M}_n$ is known to factorize into two currents as in \eref{split-fullstring},
\begin{align}
	{\cal M}^{\rm open}_n(i,\pmb A,j,\pmb B(k))\xrightarrow[]{\eref{kinematic-condition}}\epsilon_k\cdot{\cal J}^{\rm open}_{|A|+3}(i,\pmb A,j,\kappa)\times{\cal J}^{\rm open}_{|B|+3}(i_{\phi},j_{\phi},\pmb B(\kappa'_{\phi}))\,.
\end{align}
Given that the $F^3$ amplitude represents the sub-leading $\mathcal{O}(\alpha')$ term in the low-energy expansion of the open-string amplitude, one expects the following structure at linear order:
\bea
	{\cal M}_n^{{\rm open}(1)}(i,\pmb A,j,\pmb B(k))\xrightarrow[]{\eref{kinematic-condition}}&\epsilon_k\cdot{\cal J}_{|A|+3}^{{\rm open}(0)}(i,\pmb A,j,\kappa)\times{\cal J}_{|B|+3}^{{\rm open}(1)}(i_{\phi},j_{\phi},\pmb B(\kappa'_{\phi}))\nn
	&+\epsilon_k\cdot{\cal J}_{|A|+3}^{{\rm open}(1)}(i,\pmb A,j,\kappa)\times{\cal J}_{|B|+3}^{{\rm open}(0)}(i_{\phi},j_{\phi},\pmb B(\kappa'_{\phi}))\,,
\eea
where the superscript $(i)$ denotes the coefficient of the $\alpha'^i$ term. Our result in \eqref{split-F3} precisely matches this expected pattern.

In our derivation, the mixed ${\rm BAS}\oplus{\rm YM}$ currents are defined through the application of differential transmutation operators. Within this framework, we interpret $\epsilon_k\cdot{\cal J}^{F^3}_{|A|+3}$ and $\epsilon_k\cdot{\cal J}^{\rm YM}_{|A|+3}$ as the $F^3$ and YM currents, respectively, as their expansions align with those of the corresponding on-shell amplitudes. However, a technical distinction arises: as indicated in \eqref{J-F3}, the momentum $k_{\kappa}$ within the kinematic factor ${\rm Tr}({\rm F}_{\pmb\rho})$ must be replaced by the external momentum $k_k$ when expanding the $F^3$ current. In the pure YM limit, this replacement is effectively trivial because the $k_B$ components in the contraction $\epsilon_m\cdot {\rm F}_{\pmb D}\cdot\epsilon_n$ are annihilated by the kinematic conditions \eqref{kinematic-condition}.

\section{Factorizations of $R^2$ and $R^3$ amplitudes}
\label{sec-R2R3}
%%%%%%%%%%%%%%%%%%%%%%%

In this section, we investigate the $2$-split behavior of $R^2$ and $R^3$ gravitational amplitudes, which can be understood as GR amplitudes featuring single or double insertions of higher-derivative vertices. Such higher-derivative vertices arise from the low-energy effective action of bosonic closed-string theory:
\bea
S=-{2\over\kappa^2}\,\int\,d^4x\sqrt{-g}\,\Big[R-2\,(\partial_\mu\phi)^2-{1\over12}\,H^2+{\a'\over4}\,e^{-2\phi}\,G_2+\a'^2\,e^{-4\phi}\,
\big({I_1\over48}+{G_3\over24}\big)+{\cal O}(\a'^3)\Big]\,,~~\label{lowenergy-string}
\eea
where $G_2$ represents the Gauss-Bonnet term (quadratic in the Riemann tensor), while $I_1$ and $G_3$ are cubic in the Riemann tensor. This effective action implies that tree-level graviton amplitudes at $\mathcal{O}(\alpha')$ arise solely from a single insertion of $G_2$. Conversely, at $\mathcal{O}(\alpha'^2)$, contributions emerge both from single insertions of $I_1$ or $G_3$ and from double insertions of $R^2$ operators mediated by an intermediate dilaton.

Since the $R^2$ and $R^3$ operators are understood as sub-leading and sub-sub-leading corrections to Einstein-gravity amplitudes, the usual GR amplitudes considered in this section refer to purely Einstein amplitudes, rather than the full Einstein$\oplus$B-field$\oplus$dilaton system. In other words, the polarization tensor of each graviton is decomposed as $\epsilon^{\mu\nu}\equiv \epsilon^\mu\W\epsilon^\nu$, where $\epsilon^\mu=\W\epsilon^\nu$.

The expansions of these amplitudes can be constructed via the double copy approach \cite{Zhou:2024qwm}. A direct method involves exploiting the corresponding CHY integrands provided in \cite{He:2016iqi}:
\bea
{\cal I}_n^{R^2}={\cal P}_n(\epsilon)\,{\rm Pf}'{\Psi}_n(\W\epsilon)\,,~~~{\cal I}_n^{R^3}={\cal P}_n(\epsilon)\,{\cal P}_n(\W\epsilon)\,. ~~\label{F3R2-integrand}
\eea
By plugging the expansion of ${\cal P}_n$ in \eref{F3R2-integrand}, and utilizing the standard CHY integrands for mixed theories
\bea
{\cal I}^{{\rm GR}\oplus{\rm YM}}(\pmb\rho)&=&\big({\rm PT}(\pmb\rho)\,{\rm Pf}'\Psi^{\overline{\rho}}(\epsilon)\big)\,{\rm Pf}'\Psi_n(\W\epsilon)\,,\nn
{\cal I}^{{\rm GR}\oplus{\rm YM}\oplus{\rm BAS}}(\pmb\rho|\pmb\rho')&=&\big({\rm PT}(\pmb\rho)\,{\rm Pf}'\Psi^{\overline{\rho}}(\epsilon)\big)\,\big({\rm PT}(\pmb\rho')\,{\rm Pf}'\Psi^{\overline{\rho'}}(\epsilon)\big)\,,
\eea
one immediately get
\bea
{\cal A}^{R^2}_n=\sum_{\pmb\rho,2\leq|\rho|\leq n}\,{\rm Tr}({\rm F}_{\pmb\rho})\,{\cal A}^{{\rm GR}\oplus{\rm YM}}(\pmb\rho)\,.~~~~\label{expan-R^2}
\eea
and
\bea
{\cal A}^{R^3}_n&=&\sum_{\pmb\rho,2\leq|\rho|\leq n}\,\sum_{\pmb\rho',2\leq|\rho'|\leq n}\,{\rm Tr}({\rm F}_{\pmb\rho})\,{\rm Tr}({\rm \W F}_{\pmb\rho'})\,{\cal A}^{{\rm GR}\oplus{\rm YM}\oplus{\rm BAS}}_n(\pmb\rho|\pmb\rho')\,,~~~~\label{expan-R^3}
\eea
where, the mixed amplitudes ${\cal A}^{{\rm GR}\oplus{\rm YM}}$ and ${\cal A}^{{\rm GR}\oplus{\rm YM}\oplus{\rm BAS}}$ can be obtained by applying the transmutation operators to the pure gravity amplitude ${\cal A}^{\rm GR}$:
\begin{align}
	{\cal A}^{{\rm GR}_n\oplus{\rm YM}}(\pmb\rho)={\cal T}[\pmb\rho]{\cal A}^{\rm GR}_n\,,~~~{\cal A}^{{\rm GR}\oplus{\rm YM}\oplus{\rm BAS}}_n(\pmb\rho|\pmb\rho')={\cal T}[\pmb\rho]{\cal \W T}[\pmb\rho']{\cal A}^{\rm GR}_n\,.
\end{align}
$\W {\cal T}[\pmb\rho']$ and ${\rm \W F}_{\pmb\rho'}$ denote replacing $\epsilon$ in ${\cal T}[\pmb\rho']$ and  ${\rm F}_{\pmb\rho'}$ by $\W\epsilon$, respectively. Again, we have used the observation that the expansion does not affect the measure of contour integration. It is important to note that while the expansion \eqref{expan-R^2} completely recovers the string-theoretic correction at order ${\cal O}(\alpha')$, the expansion \eqref{expan-R^3} does not capture the full string correction at order ${\cal O}(\alpha'^2)$. This discrepancy arises from the double-copy structure of string amplitudes. As $F^3$ amplitudes correspond to the sub-leading open-string correction ${\cal M}^{{\rm open}(1)}$, the integrand ${\cal I}^{R^3}$ in \eref{F3R2-integrand} describes the symmetric product ${\cal M}^{{\rm open}(1)} \times {\cal M}^{{\rm open}(1)}$. However, the full closed-string amplitude ${\cal M}^{\rm closed} = {\cal M}^{\rm open} \times {\cal M}^{\rm open}$ at order ${\cal O}(\alpha'^2)$ also requires contributions from other combinations, like the product of the sub-sub-leading open-string term with the leading-order term (${\cal M}^{\text{open}(2)}_n \times {\cal M}^{\text{open}(0)}_n$).

In the rest of this section, we call amplitudes defined in \eref{expan-R^2} and \eref{expan-R^3} the $R^2$ and $R^3$ amplitudes, respectively. We find $2$-splits of $R^2$ and $R^3$ amplitudes by using the method in the previous section, and proceed by comparing these $2$-splits with the predictions from closed-string theory. As we will demonstrate, under the kinematic constraints which yield the $2$-split of GR amplitudes, these $R^2$ and $R^3$ amplitudes behave as in \eref{split-generalized}. Furthermore, the $2$-split of $R^2$ amplitudes perfectly aligns with string-theoretic expectations. However, the $R^3$ case exhibits a discrepancy. This is because the $R^3$ amplitude does not capture the full string correction at $\mathcal{O}(\alpha'^2)$, as explained above. Despite this, the $R^3$ results provide strong evidence supporting the general $2$-split structure predicted for full closed-string amplitudes.

When imposing the condition $s_{ab}=0$, a quite non-trivial problem for un-ordered graviton amplitudes is the divergences from propagators $1/s_{ab}$. Fortunately, this problem was systematically solved in \cite{Zhou:2025tvq} for the configuration $\{1,\cdots,n\}=\{i,j\}\cup A\cup B$, by showing the cancellation of divergent terms. For the current case $\{1,\cdots,n\}=\{i,j,k\}\cup A\cup B$, a special leg $k$ is moved from $B$ into $\{i,j,k\}$. This procedure eliminates divergences from $1/s_{ak}$, and does not cause any new divergence. Consequently, the remaining divergences cancel in exactly the same way, thus one need not to worry about this problem.

%%%%%%%%%%%%%%%%%%%%%%%
\subsection{$R^2$ amplitudes}
\label{subsec-R2}
%%%%%%%%%%%%%%%%%%%%%%%

As established in \eqref{expan-R^2}, any $R^2$ amplitude can be expanded into a sum of ${\rm GR}\oplus{\rm YM}$ amplitudes:
\bea
{\cal A}^{R^2}_n=\sum_{\pmb\rho,2\leq|\rho|\leq n}\,{\rm Tr}({\rm F}_{\pmb\rho})\,{\cal A}^{{\rm GR}\oplus{\rm YM}}_n(\pmb\rho)\,.\label{expan-R2}
\eea
From a string-theoretic perspective, the contribution at $\mathcal{O}(\alpha')$ arises from the product of the sub-leading open-string amplitude and the leading-order one: either ${\cal M}^{{\rm open}(1)}_n \times {\cal M}^{{\rm open}(0)}_n$ or ${\cal M}^{{\rm open}(0)}_n \times {\cal M}^{{\rm open}(1)}_n$. Since we identify $\epsilon^\mu = \tilde{\epsilon}^\mu$, these two cases are equivalent, and it is sufficient to consider the expansion in a single sector.

As discussed in the previous section, under the kinematic constraint \eqref{kinematic-condition}, the admissible choices of $\rho$ in \eqref{expan-R^2} and \eqref{expan-R^3} are: (1) $\rho=A^{\rm sub}_{ijk}$; (2) $\rho=B^{\rm sub}$; (3) $\rho=\{m,n\}\cup D\cup B^{\rm sub}$.

 The GR amplitude factorizes as \cite{Cao:2024gln, Cao:2024qpp}
\bea
{\cal A}^{\rm GR}_n\,\xrightarrow[]{\eref{kinematic-condition}}\,\epsilon_k\cdot{\cal J}^{\rm GR}_{|A|+3}\cdot\W\epsilon_k\,\times\,{\cal J}^{{\rm GR}\oplus{\rm Tr}(\phi^3)}_{|B|+3}(i_{\phi},j_{\phi},\kappa'_{\phi})\,,~~\label{split-GR}
\label{fac-GR}
\eea
where ${\cal J}^{\rm GR}$ and ${\cal J}^{{\rm GR}\oplus{\rm Tr}(\phi^3)}$ contain the external legs from the respective sets $\{i,j,\kappa\}\cup A$ and $\{i,j,\kappa'\}\cup B$. Substituting this into \eqref{expan-R2} yields
\begin{align}\label{eq:R^$2$-split}
	{\cal A}^{R^2}_n\xrightarrow[]{\eref{kinematic-condition}}\sum_{\pmb\rho,2\leq|\rho|\leq n}\,{\rm Tr}({\rm F}_{\pmb\rho})\,{\cal T}[\pmb\rho] \left[\epsilon_k\cdot{\cal J}^{\rm GR}_{|A|+3}\cdot\W\epsilon_k\,\times\,{\cal J}^{{\rm GR}\oplus{\rm Tr}(\phi^3)}_{|B|+3}(i_{\phi},j_{\phi},\kappa'_{\phi})\right]\,.
\end{align}
\textbf{Case (1)}: $\rho=A^{\rm sub}_{ijk}$\\
In this case $\rho=A^{\rm sub}_{ijk}$, ${\cal T}[\pmb\rho]$ acts exclusively on the first current ${\cal J}^{\rm GR}_{|A|+3}$ in \eqref{eq:R^$2$-split}. The resulting factorization is:
\bea
P_{(1)}\,&\xrightarrow[]{\eref{kinematic-condition}}&\,\sum_{\pmb\rho,\rho=A^{\rm sub}_{ijk}}\,{\rm Tr}({\rm F}_{\pmb\rho})\,{\cal J}^{{\rm GR}\oplus{\rm YM}}_{|A|+3}(\pmb\rho)\,\times\,{\cal J}^{{\rm GR}\oplus{\rm Tr}(\phi^3)}_{|B|+3}(i_{\phi},j_{\phi},\kappa'_{\phi})\nn
&=&\epsilon_k\cdot{\cal J}^{R^2}_{|A|+3}\cdot\W\epsilon_k\,\times\,{\cal J}^{{\rm GR}\oplus{\rm Tr}(\phi^3)}_{|B|+3}(i_{\phi},j_{\phi},\kappa'_{\phi})\,,~~\label{2split-R2-part1}
\eea
where the $R^2$ current is defined by the sum over ordered sets $\pmb\rho$ with the replacement $k_\kappa \to k_k$ in ${\rm Tr}({\rm F}_{\pmb\rho})$:
\bea
\epsilon_k\cdot{\cal J}^{R^2}_{|A|+3}\cdot\W\epsilon_k=\sum_{\pmb\rho,\rho=A^{\rm sub}_{ijk}}\,{\rm Tr}({\rm F}_{\pmb\rho})\big|_{k_\kappa\to k_k}\,{\cal J}^{{\rm GR}\oplus{\rm YM}}_{|A|+3}(\pmb\rho)\,.~~\label{defin-JR2}
\eea
We refer to \eqref{defin-JR2} as the $R^2$ current, since the expansion in \eqref{defin-JR2} becomes identical to the $R^2$ amplitude expansion \eqref{expan-R2} when the momentum of $\kappa$ is taken on-shell.\\
\textbf{Case (2)}: $\rho=B^{\rm sub}$\\
In this case, $\rho=B^{\rm sub}$, the differential operator ${\cal T}[\pmb\rho]$ acts only on the second factor ${\cal J}^{{\rm GR}\oplus{\rm Tr}(\phi^3)}_{|B|+3}$ in \eqref{eq:R^$2$-split}, yielding a double-trace current in the combined ${\rm GR}\oplus{\rm YM}\oplus{\rm BAS}$ theory:
\bea
{\cal T}[\pmb\rho]{\cal J}^{{\rm GR}\oplus{\rm Tr}(\phi^3)}_{|B|+3}(i_{\phi},j_{\phi},\kappa'_{\phi})={\cal J}^{{\rm GR}\oplus{\rm YM}\oplus{\rm BAS}}_{|B|+3}(\pmb B^{\rm sub};i,j,\kappa'|i,j,\kappa')\,.~~\label{fac-GYB}
\eea
In this context, particles in $\{i, j, k\}$ are interpreted as BAS scalars, those in $B^{\rm sub}$ as gluons, and the remainder as gravitons. It can be generated from
\bea
{\cal J}^{{\rm GR}\oplus{\rm Tr}(\phi^3)}_{|B|+3}(i,j,\kappa')={\cal J}^{{\rm GR}\oplus{\rm BAS}}_{|B|+3}(i,j,\kappa'|i,j,\kappa')\,,
\eea
by acting the operator ${\cal T}[\pmb B^{\rm sub}]$. Plugging \eref{fac-GYB} into \eqref{eq:R^$2$-split}, we get
\bea
&&P_{(2)}\,\xrightarrow[]{\eref{kinematic-condition}}\,\epsilon_k\cdot{\cal J}^{\rm GR}_{|A|+3}\cdot\W\epsilon_k\,\times\,\Big(\sum_{\pmb B^{\rm sub}}\,{\rm Tr}({\rm F}_{\pmb B^{\rm sub}})\,{\cal J}^{{\rm GR}\oplus{\rm YM}\oplus{\rm BAS}}_{|B|+3}(\pmb B^{\rm sub};i,j,\kappa'|i,j,\kappa')\Big)\,,~~\label{2split-R2-case2}
\eea
\textbf{Case (3)}: $\rho=\{m,n\}\cup D\cup B^{\rm sub}$\\
In the final case, $\rho=\{m,n\}\cup D\cup B^{\rm sub}$, we observe that the structure of $\pmb\rho$ in \eref{rhoAB} and the factorization behavior of ${\rm Tr}({\rm F}_{\pmb\rho})$ in \eref{fac-trF} still holds. Furthermore, the factorization of the combinatorial operator ${\cal T}[\pmb\rho]$ in \eref{operator} also holds. Thus we find
\bea
{\cal T}[\pmb\rho]{\cal A}^{\rm GR}={\cal J}^{{\rm GR}\oplus{\rm YM}}_{|A|+3}(m,\pmb D,n)\,\times\,{\cal J}^{{\rm GR}\oplus{\rm YM}\oplus{\rm BAS}}_{|B|+3}(\pmb B^{\rm sub}_{[ijk]}|i,j, \kappa')\,,
\eea
where ordered sets $\pmb B^{\rm sub}_{[ijk]}$ for different choices of $(m,n)$ are listed in \eref{Bijk}, and
\bea
P_{(3)}\,\xrightarrow[]{\eref{kinematic-condition}}\,\epsilon_k\cdot{\cal J}^{\rm GR}_{|A|+3}\cdot\W\epsilon_k\,\times\,\Big(\sum_{(m,n)}\,\sum_{\pmb B^{\rm sub}}\,(-)^{|B^{\rm sub}|-1}\,(k_n\cdot {\rm F}_{\pmb B^{\rm sub}}\cdot k_m)\,{\cal J}^{{\rm GR}\oplus{\rm YM}\oplus{\rm BAS}}_{|B|+3}(\pmb B^{\rm sub}_{[ijk]}|i,j,\kappa')\Big)\,,~~\label{2split-R2-case3}
\eea
where
\bea
\epsilon_k\cdot{\cal J}^{\rm GR}_{|A|+3}\cdot\W\epsilon_k=\sum_{\pmb D}\,(-)^{|D|}\,(\epsilon_m\cdot {\rm F}_{\pmb D}\cdot\epsilon_n)\,\cdot{\cal J}^{{\rm GR}\oplus{\rm YM}}_{|A|+3}(m,\pmb D,n)\,.
\eea

By combining the contributions from all three cases, we arrive at the complete factorization formula for the $R^2$ amplitude:
\bea
{\cal A}^{R^2}_n&\xrightarrow[]{\eref{kinematic-condition}}&\epsilon_k\cdot{\cal J}^{R^2}_{|A|+3}\cdot\W\epsilon_k\,\times\,
{\cal J}^{{\rm Tr}(\phi^3)\oplus{\rm GR}}_{|B|+3}(i_\phi,j_\phi,\kappa'_\phi)\nn
&&+\epsilon_k\cdot{\cal J}^{\rm GR}_{|A|+3}\cdot\W\epsilon_k\,\times\,{\cal J}^Y_{|B|+3}(i_\phi,j_\phi,\kappa'_\phi)\,,~~\label{split-R2}
\eea
where the composite current ${\cal J}^Y$ is defined as:
\bea
{\cal J}^Y_{|B|+3}(i_\phi,j_\phi,\kappa'_\phi)&=&\sum_{\pmb B^{\rm sub}}\,{\rm Tr}({\rm F}_{\pmb B^{\rm sub}})\,{\cal J}^{{\rm GR}\oplus{\rm YM}\oplus{\rm BAS}}_{|B|+3}(\pmb B^{\rm sub};i,j,\kappa'|i,j,\kappa')\nn
&&
+\sum_{(m,n)}\,\sum_{\pmb B^{\rm sub}}\,(-)^{|B^{\rm sub}|-1}\,(k_n\cdot {\rm F}_{\pmb B^{\rm sub}}\cdot k_m)\,{\cal J}^{{\rm GR}\oplus{\rm YM}\oplus{\rm BAS}}_{|B|+3}(\pmb B^{\rm sub}_{[ijk]}|i,j,\kappa')\,.~~\label{JY}
\eea
Specifically, the sub-leading order ($\alpha'$) expansion of the closed-string amplitude in \eref{split-fullstring} is expected to follow the structure:
\bea
{\cal M}^{{\rm closed}(1)}_n\xrightarrow[]{\eref{kinematic-condition}}\epsilon_k\cdot{\cal J}^{{\rm closed}(1)}_{|A|+3}\cdot\W\epsilon_k\times{\cal J}^{{\rm closed}(0)}_{|B|+3}(i_\phi,j_\phi,\kappa'_\phi)
+\epsilon_k\cdot{\cal J}^{{\rm closed}(0)}_{|A|+3}\cdot\W\epsilon_k\times{\cal J}^{{\rm closed}(1)}_{|B|+3}(i_\phi,j_\phi,\kappa'_\phi),
\eea
where
\bea
\epsilon_k\cdot{\cal J}^{{\rm closed}(1)}_{|A|+3}\cdot\W\epsilon_k=2\,\epsilon_k\cdot{\cal J}^{R^2}_{|A+3|}\cdot\W\epsilon_k\,,~~
{\cal J}^{{\rm closed}(1)}_{|B|+3}(i_\phi,j_\phi,\kappa'_\phi)=2\,{\cal J}^Y_{|B|+3}(i_\phi,j_\phi,\kappa'_\phi)\,,~~\label{J-sub-string}
\eea
since the $2$-split of standard GR amplitude in \eref{split-GR} forces
\bea
\epsilon_k\cdot{\cal J}^{{\rm closed}(0)}_{|A|+3}\cdot\W\epsilon_k=\epsilon_k\cdot{\cal J}^{\rm GR}_{|A|+3}\cdot\W\epsilon_k\,, ~~~{\cal J}^{{\rm closed}(0)}_{|B|+3}(i_\phi,j_\phi,\kappa'_\phi)={\cal J}^{{\rm GR}\oplus{\rm Tr}(\phi^3)}_{|B|+3}(i_{\phi},j_{\phi},\kappa'_{\phi})\,.
\eea
The factor of $2$ in \eqref{J-sub-string} originates from the double-copy structure discussed at the beginning of this section. Since the full subleading closed-string amplitude receives equivalent contributions from both the ${\cal M}^{{\rm open}(1)}_n\times{\cal M}^{{\rm open}(0)}_n$ and ${\cal M}^{{\rm open}(0)}_n\times{\cal M}^{{\rm open}(1)}_n$ sectors, the $R^2$ result (which considers a single sector) represents exactly half of the full string-theoretic correction.

%%%%%%%%%%%%%%%%%%%%%%%
\subsection{$R^3$ amplitudes}
\label{subsec-R3}
%%%%%%%%%%%%%%%%%%%%%%%

The treatment for the $R^3$ case, involving ${\rm Tr}({\rm F}_{\pmb\rho})$ and ${\rm Tr}(\W{\rm F}_{\pmb\rho'})$, is almost the same as before. Repeating the process, we decompose the amplitude into four parts based on the partitioning of the operators:
\bea
&&P_{(1)\times(1)}\,\xrightarrow[]{\eref{kinematic-condition}}\,\epsilon_k\cdot{\cal J}^{R^3}_{|A|+3}\cdot\W\epsilon_k\,\times\,
{\cal J}^{{\rm Tr}(\phi^3)\oplus{\rm GR}}_{|B|+3}(i_\phi,j_\phi,\kappa'_\phi)\,,\nn
&&P_{(2)(3)\times(2)(3)}\,\xrightarrow[]{\eref{kinematic-condition}}\,\epsilon_k\cdot{\cal J}^{\rm GR}_{|A|+3}\cdot\W\epsilon_k\,\times\,{\cal J}^Z_{|B|+3}(i_\phi,j_\phi,\kappa'_\phi)\,,\nn
&&P_{(1)\times(2)(3)}\,\xrightarrow[]{\eref{kinematic-condition}}\,\epsilon_k\cdot{\cal J}^{R^2}_{|A|+3}\cdot\W\epsilon_k\,\times\,{\cal J}^Y_{|B|+3}(i_\phi,j_\phi,\kappa'_\phi)\,,\nn
&&P_{(2)(3)\times(1)}\,\xrightarrow[]{\eref{kinematic-condition}}\,\epsilon_k\cdot{\cal J}^{R^2}_{|A|+3}\cdot\W\epsilon_k\,\times\,{\cal J}^Y_{|B|+3}(i_\phi,j_\phi,\kappa'_\phi)\,.~~\label{R3-cases}
\eea
These four parts correspond to
\bea
&&P_{(1)\times(1)}\,\sim\,{\rm case\,(1)}\,\times\,{\rm case\,(1)}\,,~~~~P_{(2)(3)\times(2)(3)}\,\sim\,{\rm case\,(2),(3)}\,\times\,{\rm case\,(2),(3)}\,,\nn
&&P_{(1)\times(2)(3)}\,\sim\,{\rm case\,(1)}\,\times\,{\rm case\,(2),(3)}\,,~~~~P_{(2)(3)\times(1)}\,\sim\,{\rm case\,(2),(3)}\,\times\,{\rm case\,(1)}\,,
\eea
respectively. In \eref{R3-cases}, the $R^2$ current $\epsilon_k\cdot{\cal J}^{R^2}_{|A|+3}\cdot\W\epsilon_k$ is defined in \eref{defin-JR2}, while the $R^3$ current $\epsilon_k\cdot{\cal J}^{R^3}_{|A|+3}\cdot\W\epsilon_k$ is defined as
\bea
\epsilon_k\cdot{\cal J}^{R^3}_{|A|+3}\cdot\W\epsilon_k=\sum_{\pmb\rho,\rho=A^{\rm sub}_{ijk}}\,\sum_{\pmb\rho',\rho'=A^{\rm sub}_{ijk}}\,{\rm Tr}({\rm F}_{\pmb\rho})\big|_{k_\kappa\to k_k}\,{\rm Tr}({\rm \W F}_{\pmb\rho'})\big|_{k_\kappa\to k_k}\,{\cal J}^{{\rm GR}\oplus{\rm YM}\oplus{\rm BAS}}_{|A|+3}(\pmb\rho|\pmb\rho')\,,~~\label{define-JR3}
\eea
according to the expansion of on-shell $R^3$ amplitudes in \eref{expan-R^3}, where the mixed currents ${\cal J}^{{\rm GR}\oplus{\rm YM}\oplus{\rm BAS}}_{|A|+3}$ are generated from ${\cal J}^{\rm GR}_{|A|+3}$ by acting differential operators.
Meanwhile, the mixed current ${\cal J}^Y_{|B|+3}$ is given in \eref{JY}, while
\bea
{\cal J}^Z_{|B|+3}&=&\sum_{\pmb B^{\rm sub}}\,\sum_{\pmb{\W B}^{\rm sub}}\,{\rm Tr}({\rm F}_{\pmb B^{\rm sub}})\,{\rm Tr}({\rm\W F}_{\pmb{\W B}^{\rm sub}})\,{\cal J}^{{\rm GR}\oplus{\rm YM}\oplus{\rm BAS}}_{|B|+3}(\pmb B^{\rm sub};i,j,\kappa'|\pmb{\W B}^{\rm sub};i,j,\kappa')\nn
&&
+\sum_{(m,n)}\,\sum_{(\W m,\W n)}\,\sum_{\pmb B^{\rm sub}}\,\sum_{\pmb{\W B}^{\rm sub}}\,(-)^{|B^{\rm sub}|+|\W B^{\rm sub}|}\,(k_n\cdot {\rm F}_{\pmb B^{\rm sub}}\cdot k_m)\,(k_{\W b}\cdot {\rm\W F}_{\pmb{\W B}^{\rm sub}}\cdot k_{\W a})\,{\cal J}^{{\rm GR}\oplus{\rm YM}\oplus{\rm BAS}}_{|B|+3}(\pmb B^{\rm sub}_{[ijk]}|\pmb{\W B}^{\rm sub}_{[ijk]})\nn
&&
+2\,\sum_{(m,n)}\,\sum_{\pmb B^{\rm sub}}\,\sum_{\pmb{\W B}^{\rm sub}}\,(-)^{|B^{\rm sub}|}\,(k_n\cdot {\rm F}_{\pmb B^{\rm sub}}\cdot k_m)\,{\rm Tr}({\rm\W F}_{\pmb{\W B}^{\rm sub}})\,{\cal J}^{{\rm GR}\oplus{\rm YM}\oplus{\rm BAS}}_{|B|+3}(\pmb B^{\rm sub}_{[ijk]}|\pmb{\W B}^{\rm sub};i,j,\kappa')\,.~~\label{define-JZ}
\eea
As can be seen, $P_{(1)\times(2)(3)}$ and $P_{(2)(3)\times(1)}$ in \eref{R3-cases} are equal to each other. This stems from the fact in pure Einstein gravity, $\epsilon_\ell=\W\epsilon_\ell$, so exchanging $\epsilon_\ell$ and $\W\epsilon_\ell$ in the definitions of ${\cal J}^{R^2}_{|A|+3}$ and ${\cal J}^Y_{|B|+3}$ has no effect. This is also the reason why the last line of ${\cal J}^Z_{|B|+3}$ in \eref{define-JZ} carries a factor $2$.

Combining four parts together, we get
\bea
{\cal A}^{R^3}_n&\xrightarrow[]{\eref{kinematic-condition}}&\epsilon_k\cdot{\cal J}^{R^3}_{|A|+3}\cdot\W\epsilon_k\,\times\,
{\cal J}^{{\rm Tr}(\phi^3)\oplus{\rm GR}}_{|B|+3}(i_\phi,j_\phi,\kappa'_\phi)\nn
&&+\epsilon_k\cdot{\cal J}^{\rm GR}_{|A|+3}\cdot\W\epsilon_k\,\times\,{\cal J}^Z_{|B|+3}(i_\phi,j_\phi,\kappa'_\phi)\nn
&&+2\,\epsilon_k\cdot{\cal J}^{R^2}_{|A|+3}\cdot\W\epsilon_k\,\times\,{\cal J}^Y_{|B|+3}(i_\phi,j_\phi,\kappa'_\phi)\,.~~\label{split-R3}
\eea
The $2$-split factorization of closed-string amplitudes in \eref{split-fullstring} implies the following structure at order $\alpha'^2$:
\bea
{\cal M}_n^{{\rm closed}(2)}&\xrightarrow[]{\eref{kinematic-condition}}&\epsilon_k\cdot{\cal J}_{|A|+3}^{{\rm closed}(2)}\cdot\W\epsilon_k\,\times\,{\cal J}_{|B|+3}^{{\rm closed}(0)}(i_\phi,j_\phi,\kappa'_\phi)\,+\,\epsilon_k\cdot{\cal J}_{|A|+3}^{{\rm closed}(0)}\cdot\W\epsilon_k\,\times\,{\cal J}_{|B|+3}^{{\rm closed}(2)}(i_\phi,j_\phi,\kappa'_\phi)\nn
&&+\,\epsilon_k\cdot{\cal J}_{|A|+3}^{{\rm closed}(1)}\cdot\W\epsilon_k\,\times\,{\cal J}_{|B|+3}^{{\rm closed}(1)}(i_\phi,j_\phi,\kappa'_\phi)\,.~~\label{fac-R3-string}
\eea
In the previous subsection, we identified the $\mathcal{O}(\alpha')$ string currents as:
\bea
\epsilon_k\cdot{\cal J}_{|A|+3}^{{\rm closed}(1)}\cdot\W\epsilon_k=2\,\epsilon_k\cdot{\cal J}^{R^2}_{|A|+3}\cdot\W\epsilon_k\,,~~~~{\cal J}_{|B|+3}^{{\rm closed}(1)}(i_\phi,j_\phi,\kappa'_\phi)=2\,{\cal J}^Y_{|B|+3}(i_\phi,j_\phi,\kappa'_\phi)\,.
\eea
Comparing these with our result in \eqref{split-R3}, it is evident that the $R^3$ factorization does not restore the complete prediction of the closed-string $2$-split.

As discussed at the beginning of this section, this discrepancy arises because the full sub-sub-leading term of the closed-string amplitude requires additional contributions from sectors such as ${\cal M}^{{\rm open}(2)}_n \times {\cal M}^{{\rm open}(0)}_n$. Nevertheless, a direct correspondence can be established: the terms $\epsilon_k \cdot {\cal J}^{R^3}_{|A+3|} \cdot \tilde{\epsilon}_k$ and ${\cal J}^Z_{|B|+3}$ in \eqref{split-R3} represent specific contributions to the string currents $\epsilon_k \cdot {\cal J}_{|A|+3}^{(2)} \cdot \tilde{\epsilon}_k$ and ${\cal J}_{|B|+3}^{(2)}$ in \eqref{fac-R3-string}, respectively. Furthermore, the final term in \eqref{split-R3} maps directly to the last term in \eref{fac-R3-string}. In this sense, we conclude that our $R^3$ amplitude defined in \eqref{expan-R^3}---which represents a specific sectoral contribution to the closed-string amplitude at order ${\cal O}(\alpha'^2)$---is entirely consistent with the expected $2$-split behavior of the full string amplitude.

%%%%%%%%%%%%%%%%%%%%%
\section{Summary and discussion}
\label{sec-conclu}
%%%%%%%%%%%%%%%%%%%%%%

In this paper, by exploiting universal expansion formulas, we have demonstrated that YM and GR amplitudes featuring specific higher-derivative vertex insertions---which are also known to exhibit hidden zeros---obey the generalized $2$-split behavior defined in \eqref{split-generalized}. Furthermore, we have shown that our results are entirely consistent with the $2$-split structures observed in full string amplitudes. Our findings imply that an amplitude exhibiting hidden zero behavior does not necessarily satisfy the standard $2$-split form of \eqref{split-phi3}. However, it remains plausible that the hidden zero phenomenon is always accompanied by a $2$-split behavior, provided one adopts the more general version presented in \eqref{split-generalized}. Consequently, it would be highly valuable to investigate whether this conjecture holds universally and to uncover the underlying physical principles that guarantee this connection.

In the generalized $2$-split formulas \eqref{split-F3}, \eqref{split-R2}, and \eqref{split-R3}, the emergence of the complex currents ${\cal J}^X$, ${\cal J}^Y$, and ${\cal J}^Z$ is not immediately intuitive. While these currents are formally defined through expansion formulas, their physical interpretation remains non-manifest and opaque. To achieve a more transparent understanding of these objects, it would be advantageous to derive the splitting behaviors using alternative frameworks that have successfully characterized $2$-splits in standard YM and GR amplitudes, such as the CHY formalism and BCFW recursion relation. The former could allow one to identify the precise integrands corresponding to these currents, while the latter could reveal the lower-point building blocks of these currents. By utilizing these ingredients, we may move beyond purely formal expansions and gain a more physical grasp of the internal structure of these higher-derivative currents.

As investigated in \cite{Arkani-Hamed:2024fyd} and \cite{Zhou:2025xly}, the $2$-split behaviors of scattering amplitudes are intimately related to their universal soft limits. More explicitly, the kinematic constraint $s_{ab}=0$ for all $a \in A, b \in B$ can be partially realized by taking the soft limit for all particles in the set $A$.\footnote{The soft limit requires the additional constraints $s_{\ell a}=0$ with $\ell\in\{i,j,k\}$, which means that the soft behavior is not strictly equivalent to the $2$-split behavior.} The resulting current ${\cal J}_{|A|+3}$ is related to the corresponding soft factor. This connection raises a compelling question regarding higher-derivative theories. For $F^3$, $R^2$, and $R^3$ amplitudes, we have shown that the $2$-split behavior is partitioned into multiple sectors, as expressed in the generalized formula \eqref{split-generalized}. One might naturally expect this structural decomposition to manifest in the soft factors as well; specifically, that the soft factors for these amplitudes are composed of a combination of multiple terms. For instance, the $F^3$ soft factor might take the form $S^{F^3} = S_1^{F^3} + S_2^{F^3}$. Investigating this conjecture and determining whether these individual soft components correspond to the distinct current contributions identified in our $2$-split analysis would be a interesting direction for future research.

\section*{Acknowledgments}

We would like to thank Prof. Bo Feng for valuable discussions and suggestions. K.Z is supported by NSFC under Grant No. 11805163.

\bibliographystyle{JHEP}

\bibliography{reference}

\end{document}